\documentclass{andp2012}
\usepackage[english]{babel}

\newcommand{\PC}[1]{\ensuremath{\left(#1\right)}}

\usepackage{kantlipsum,widetext}
\usepackage{xcolor}
\usepackage{subfig}
\title{Effects of a uniform magnetic field on twisted graphene nanoribbons}
\author[C.\,C. Soares]{Camila C. Soares\inst{1}}
\author[A.\,E.\,Obispo]{Angel E. Obispo\inst{2,3,}\footnote{Corresponding author\quad E-mail:\\~\textsf{aobispo@utp.edu.pe, angel.obispo@upn.pe}}}
\author[A.\,G.\,J.\,Vicente]{Andrés G. Jirón Vicente\inst{4}}
\author[L.\,B. Castro]{Luis B. Castro\inst{5}}
\address[1]{Departamento de Ensino, Instituto Federal do Maranh\~{a}o (IFMA), Campus Z\'{e} Doca, 65365-000, Z\'{e} Doca, MA, Brazil.}
\address[2]{Universidade Tecnológica del Perú (UTP), Lima, Perú}
\address[3]{Departamento de Ciencias, Universidad Privada del Norte (UPN), Lima, Perú.}
\address[4]{Facultad de Ciencias Naturales y Matemáticas (FCNM), Universidad Nacional del Callao (UNAC), Campus Central, 07001, Bellavista, Callao, Perú.}
\address[5]{Departamento de F\'{\i}sica, Universidade Federal do Maranh\~{a}o, Campus Universit\'{a}rio do Bacanga, 65080-805, S\~{a}o Lu\'{\i}s, MA, Brazil.}
\shortauthors{Camila C. Soares et al.}
\begin{abstract}
In the present work, the relativistic quantum motion of massless fermions in a helicoidal graphene nanoribbon under the influence of a uniform magnetic field is investigated. Considering a uniform magnetic field ($B$) aligned along the axis of helicoid, this problem is explored in the context of Dirac equation in a curved space-time. As this system does not support exact solutions due to considered background, the bound-state solutions and local density of state (LDOS) are obtained numerically by means of the Numerov method. The combined effects of width of the nanoribbon ($D$), length of ribbon ($L$), twist parameter ($\omega$) and $B$ on the equations of motion and local density of states (LDOS) are analyzed and discussed. It is verified that the presence of $B$ produces a constant minimum value of local density of state on the axis of helicoid, which is possible only for values large enough of $\omega$, in contrast to the case for $B=0$ already studied in the literature.
\end{abstract}
\shortabstract
\begin{document}
\maketitle

\section{Introduction}
\label{intro}

In the last three decades, there has been a great deal of interest in the implementation of models of general relativity in systems typically related to experimental areas, such as Relativistic cosmology \cite{ellis2012relativistic}, Orbital dynamic \cite{steinhoff2016dynamical,iorio2012constraining}, and most recently, condensed matter Physics \cite{cortijo2012geometrical,cortijo2007electronic,3,atanasov2015helicoidal}. Particularly, the latter has become the subject of a large number of experimental and theoretical works \cite{cortijo2012geometrical} that employ gravitational models to attempt to describe nanostructures with novel and complicated geometries. Some relevant examples include cosmic strings, which have been compared to disclinations in liquid crystals \cite{cortijo2007effects,cortijo2007cosmological}, black-hole space-times in ion rings \cite{horstmann2010hawking}, curved Lorentzian (pseudo–Riemannian) geometry to study the acoustic propagation in classical fluids \cite{visser1993acoustic}, 2D Weitzenbock geometry to represent variations of the hopping parameters in graphene \cite{volovik2014,zubkov2015}, a effective torsion and magnetic field induced by dislocations in strain-graphene \cite{volovik2015}, and de Sitter space-time for Bose-Einstein condensates \cite{fedichev2003gibbons}. Here, we will focus our attention on another special type of surface, the so-called helicoidal nanoribbon.

The helicoidal surface is one of the simplest periodic curved surfaces in one direction \cite{hoffman2002deforming} and that has vanished mean curvature (minimal surfaces). Helicoidal microstructures occur often in biology, for example, in macro-molecules as DNA \cite{barbi1999helicoidal,dang2012nonlinear,dauxois1991dynamics} and in the animal and plant kingdoms \cite{wilts2014natural}. On the other hand, condensed matter examples include screw dislocations in liquid crystals \cite{kamien1999minimal,kamien2001order} and certain ferroelectric crystals \cite{walba2000ferroelectric}, where the effects of curvature led to the emergence of unusual behavior of charge carriers in strong external electric and magnetic fields.

In the context of graphene, some theoretical investigations has shown that the dynamics of low-energy free electrons on helicoidal surfaces is affected by an effective potential induced by curvature effects in (2+1)-dimensional curved space-time, in Schr\"{o}dinger \cite{5} and Dirac \cite{6,atanasov2015helicoidal} descriptions. In addition, we can also mention some important examples related to the combined effects of the helicoidal geometry and external background fields, for example, particles confined on a helicoidal graphene ribbon interacting with da Costa potential \cite{3}, the emergency of pseudo-Landau levels due to a strain-induced pseudomagnetic field \cite{zhang2014strain}, or helicoidal graphene in presence of two specific magnetic field configurations \cite{17}.

Inspired by these works, our work proposes to explore via numerical calculations the dynamics in (2+1)-dimensional massless Dirac particles on a helicoidal surface in the presence of a uniform magnetic field aligned along the axis of helicoid. We analyze the behavior of the effective potentials for some values of the magnetic field ($B$) and the twist parameter ($\omega$). Also, we calculate the local density of states (LDOS) as a point of convergence between general relativity and condensed matter Physics, similar to performed by Watanabe et al. in \cite{6}, where a partial local density of states was built in the context of a graphene lattice. From this study, an unexpected and counterintuitive behavior of LDOS is found. Finally, we conclude that this novel result is a consequence of the choice in the configuration of the magnetic field and a possible connection with condensed matter Physics is qualitatively discussed.

\section{Massless Dirac fermions in a helicoidal graphene nanoribbon}
\label{sec:1}
It is widely known that low-energy electronic excitations in a clean flat graphene are well described by massless two dimensional Dirac equation. Nevertheless, if we want to include non-trivial intrinsic curvature effects, it is necessary to extend this Dirac formulation for graphene to its curved space version. For this purpose, we consider the helicoidal graphene nanoribbon as a continuous structure without any distortion or strain, where the discreteness of the hexagonal lattice or the variations of the hopping parameters are not taken into account. Based on these considerations, we have that low-energy electronic excitations in the helicoidal graphene nanoribbon can be performed by the following bidimensional Dirac equation in curved Riemann space \cite{6,atanasov2015helicoidal} ($\hbar=c=1$)%
\begin{equation}\label{dkp}
 i\gamma ^{\mu }\nabla_{\mu} \Psi =0\,,\quad (\mu=0,1,2)
\end{equation}%
\noindent where the covariant derivative
\begin{equation}\label{der_cov}
\nabla_{\mu}=\partial _{\mu }+\Gamma_{\mu}\,,
\end{equation}
\noindent and $\Psi$ represents a two-component spinor. The affine connection is defined by
\begin{equation}\label{affine}
\Gamma_{\mu}=\frac{1}{8}\,\omega_{\mu (a)(b)}[\gamma^{(a)},\gamma^{(b)}]\,.
\end{equation}
\noindent The curved-space gamma matrices are
\begin{equation}\label{beta_curved}
\gamma ^{\mu }=e^{\mu}\,_{(a)}\,\gamma^{(a)}\,,
\end{equation}
\noindent and satisfy the algebra  $\left\{ \gamma ^{\mu
},\gamma ^{\nu }\right\} =2g^{\mu \nu }\mathbf{I}_{2\times 2}~$, where $g^{\mu \nu }$ is the metric tensor.
The \textit{tetrads} $e_{\mu}\,^{(a)}(x)$ satisfy the relations%
\begin{eqnarray}
\eta^{(a)(b)} &=& e_{\mu}\,^{(a)}\,e_{\nu}\,^{(b)}\,g^{\mu\nu}\,,\label{tetr1}\\
g_{\mu\nu} &=& e_{\mu}\,^{(a)}\,e_{\nu}\,^{(b)}\,\eta_{(a)(b)}\,,\label{tetr2}
\end{eqnarray}
\noindent and
\begin{equation}\label{tetr3}
e_{\mu}\,^{(a)}\,e^{\mu}\,_{(b)}=\delta^{(a)}_{(b)}\,,
\end{equation}%
\noindent the Latin indexes being raised and lowered by the Min\-kowski metric tensor $\eta^{(a)(b)}$ with signature $(+,-,-)$ and the Greek ones by the metric tensor $g^{\mu\nu}$.

\noindent The spin connection $\omega_{\mu (a)(b)}$ is given by
\begin{equation}\label{con}
\omega_{\mu}\,^{(a)(b)}=e_{\alpha}\,^{(a)}\,e^{\nu (b)}\,\Gamma_{\mu\nu}^{\alpha}
-e^{\nu (b)}\partial_{\mu}e_{\nu}\,^{(a)}
\end{equation}%
\noindent with $\omega_{\mu}\,^{(a)(b)}=-\omega_{\mu}\,^{(b)(a)}$ and $\Gamma_{\mu\nu}^{\alpha}$ are the Christoffel symbols given by
\begin{equation}\label{symc}
\Gamma_{\mu\nu}^{\alpha}=\frac{g^{\alpha\beta}}{2}\left( \partial_{\mu}g_{\beta\nu}+
\partial_{\nu}g_{\beta\mu}-\partial_{\beta}g_{\mu\nu} \right).
\end{equation}%

\section{Geometric setup of helicoidal surfaces}
\label{sec:2}

\begin{figure}[t]
\subfloat[]{\label{fig_w1}\includegraphics[scale=0.3]{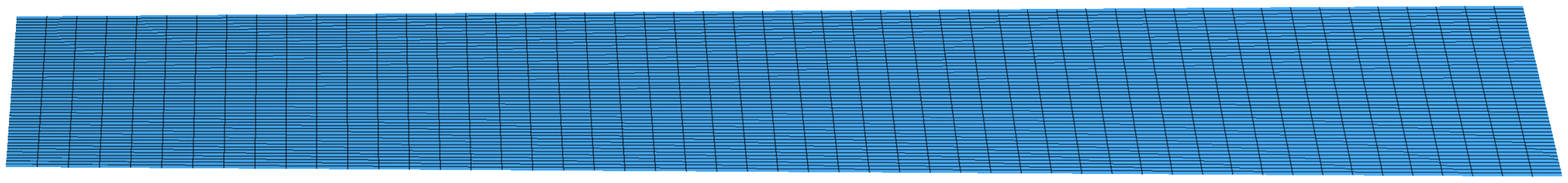}}\\
\subfloat[]{\label{fig_w2}\includegraphics[scale=0.3]{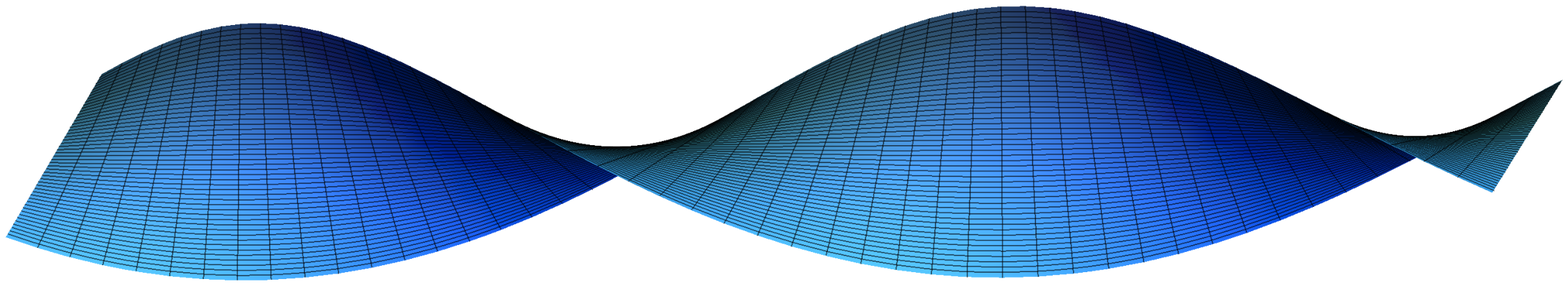}}\\
\subfloat[]{\label{fig_w3}\includegraphics[scale=0.3]{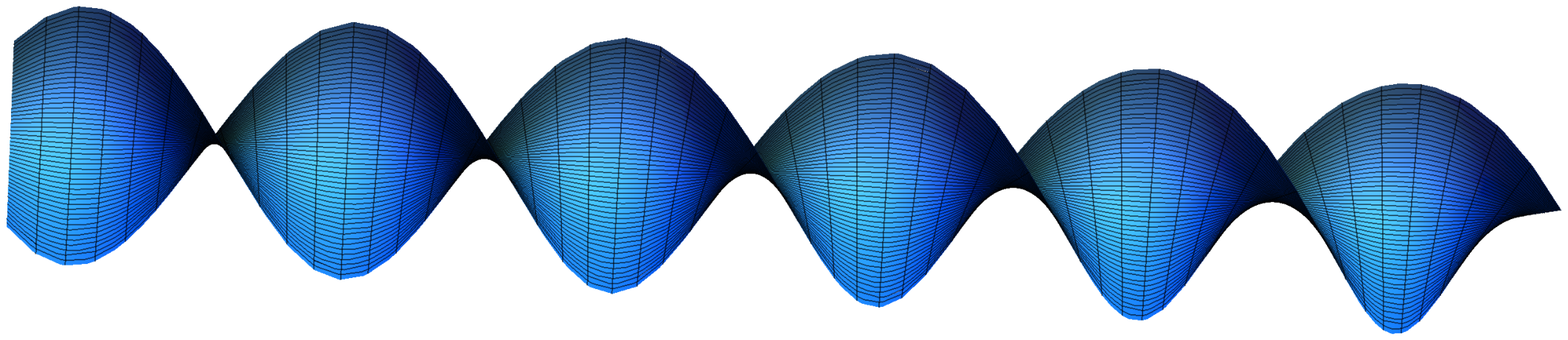}}
\caption{Helicoidal graphene nanoribbon profiles for (a) $\omega=0.01$, (b) $\omega=0.5$ and (c) $\omega=1.5$, considering $L=10\pi$.}
\label{fig_w}
\end{figure}

The helicoidal background geometry is described using the following parametrization
\begin{eqnarray}
x &=&v,  \notag \\
y &=&u\cos \left( \omega v\right) ,  \label{1} \\
z &=&u\sin \left( \omega v\right) ,  \notag
\end{eqnarray}%
\noindent where $v\in \left[ -\frac{L}{2},\frac{L}{2}\right]$ and $u\in \left[ -\frac{%
D}{2},\frac{D}{2}\right] $. Here $D$ is the width of the nanoribbon and $L$ is the total length of the ribbon which is
aligned around the $x$-axis. The constant $\omega =\frac{2\pi m}{L}$ is a real number that determines the
chirality of the surface (twist parameter), and $m$ is the number of \ $2\pi$ twists. In \ref{fig_w}, the helicoidal graphene nanoribbon profiles for $\omega=0.01$, $\omega=0.5$ and $\omega=1.5$ are shown. One can see that the twist of the nanoribbon increases with the value of $\omega$. Also is verified that as $\omega\rightarrow 0$, the flat graphene is reproduced. \ With this parameterization, the helicoidal surface can be mapped into the $(2+1)$%
-dimensional space-time by the following line element,%
\begin{equation}
ds^{2} =dt^{2}-du^{2}-f(u) dv ^{2},  \label{metric}
\end{equation}%
\noindent where $f(u)=1+\omega ^{2}u^{2}$, and the temporal coordinate $t$ was projected trivially from flat $(3+1)$%
-dimensional space-time where the helicoidal ribbon lives. This allows us read
the metric components on the helicoidal surface directly from (\ref{metric}) as%
\begin{equation}
g_{\alpha \beta }=\left(
\begin{array}{ccc}
1 & 0 & 0 \\
0 & -1 & 0 \\
0 & 0 & -f(u)%
\end{array}%
\right) ,  \label{metric matrix}
\end{equation}%

The basis tetrad $e^{\mu}\,_{(a)}$ from the line element (\ref{metric}) is chosen to be
\begin{equation}\label{tetra}
e^{\mu}\,_{(a)}=%
\begin{pmatrix}
1 & 0 & 0\\
0 & 1 & 0\\
0 & 0 & \frac{1}{\sqrt{f(u)}}
\end{pmatrix}%
\,.
\end{equation}%
\noindent For the specific basis tetrad (\ref{tetra}) the curved-space gamma matrices read
\begin{eqnarray}
\gamma^{t} &=& \gamma^{(0)}\,,  \label{betat} \\
\gamma^{u} &=& \gamma^{(1)}\,,   \label{betar} \\
\gamma^{v} &=& \frac{\gamma ^{(2)}}{\sqrt{f(u)} }\,,  \label{betaphi}
\end{eqnarray}%
\noindent the non-zero components of the Christoffel symbols are given by
\begin{eqnarray}
\Gamma _{vv}^{u} &=& -u\omega ^{2},  \label{3.3a} \\
\Gamma _{uv}^{v} &=& \Gamma _{vu}^{v}=\frac{u\omega ^{2}}{f(u)}\,,\label{3.3b}
\end{eqnarray}%
\noindent the non-zero spin connection is given by
\begin{equation}
\omega _{v(1)(2)} =-\omega _{v(2)(1)} =\frac{\omega ^{2}u}{\sqrt{f(u)}}\,,\label{3.3d}
\end{equation}
\noindent and the non-zero affine connection is
\begin{equation}\label{gphi}
\Gamma_{v}=\frac{1}{4}\frac{\omega ^{2}u}{\sqrt{f(u)}}[\gamma^{(1)},\gamma^{(2)}]\,.
\end{equation}
\noindent We choose the representation which one,
\begin{equation}
\gamma ^{(0)}=\sigma ^{(3)}=\begin{pmatrix}
1 &  0\\
0 & -1
\end{pmatrix}\,,
\end{equation}
\begin{equation}
\gamma ^{(1)}=i\sigma ^{(2)}=\begin{pmatrix}
0 &  1\\
-1 & 0
\end{pmatrix}\,,
\end{equation}
\noindent and
\begin{equation}
\gamma ^{(2)}=-i\sigma ^{(1)}=\begin{pmatrix}
0 &  -i\\
-i & 0
\end{pmatrix}\,.
\end{equation}
\noindent Thereby, in this representation the covariant derivative gets
\begin{eqnarray}
\nabla_{t} &=& \partial_{0}\,,\label{dt}\\
\nabla_{u} &=& \partial_{u}\,,\label{dr}\\
\nabla_{v} &=& \partial_{v}-\frac{\omega ^{2}u}{\sqrt{f(u)}}\frac{i\sigma ^{(3)}}{2}\,.\label{dphi}
\end{eqnarray}
\noindent Having set up the Dirac equation in a curved space-time, we are now in a position to use the machinery developed above in order to solve the Dirac equation in this background with some specific forms for external interactions.

\section{Massless fermions in a helicoidal graphene nanoribbon in the presence of an electromagnetic field}
\label{sec:3}

In this section, we concentrate our efforts in the study of the electromagnetic interaction embedded in the background of a helicoidal nanoribbon. For this external interaction we use the minimal coupling ($e=1$)
\begin{equation}
\partial_{\mu}\rightarrow \partial_{\mu}+iA_{\mu}\,.
\end{equation}
\noindent Considering only the $v$-component $A_{\mu}=(0,0,A_{v}(u))$, the Dirac equation (\ref{dkp}) becomes
\begin{equation}
\left( i\gamma ^{\mu }\nabla_{\mu }-\gamma^{v}A_{v}\right) \Psi =0.
\label{Dirac}
\end{equation}%
\noindent As the interaction is time-independent one can write
\begin{equation}\label{5}
\Psi (u,v)=e^{-iEt}\left(
\begin{array}{c}
\Psi ^{+}(u,v) \\
\Psi ^{-}(u,v)%
\end{array}%
\right)\,,
\end{equation}%
\noindent where $E$ is the energy of the fermion, $\Psi^{+}$ and $\Psi^{-}$ are the upper and lower components, respectively. Inserting (\ref{betat}), (\ref{betar}), (\ref{betaphi}), (\ref{dt}), (\ref{dr}), (\ref{dphi}) and (\ref{5}) in (\ref{Dirac}), we obtain two coupled first-order equations for the upper ($\Psi^{+}$) and lower ($\Psi^{-}$) components
\begin{eqnarray}
\left( i\partial _{u}+\frac{i\omega ^{2}u}{2 f(u)}+\frac{1}{\sqrt{f(u)}}\partial _{v}-iA_{v}\right)
\Psi ^{-} &=&-E\Psi ^{+}, \label{6.1} \\
\left( i\partial _{u}+\frac{i\omega ^{2}u}{2 f(u)}-\frac{1}{\sqrt{f(u)}}\partial _{v}+iA_{v}\right)
\Psi ^{+} &=&-E\Psi ^{-}. \label{6.2}
\end{eqnarray}%
\noindent Since $v$ is a cyclic coordinate, we can consider $\omega v$ and $L_{v}=-%
\frac{i}{\omega }\frac{\partial }{\partial v}$ as the azimuthal angle and the
angular momentum operator around the axis of the helicoid (cylindrical
symmetry), respectively. The operator $L_{v}$ satisfies $L_{v}\Psi (u,v)=l\Psi (u,v)$, with $l~\epsilon ~%
\mathbb{Z}
$ and it commutes with the Hamiltonian
\begin{equation}
H=\gamma ^{(0)}\gamma ^{i}\left( \widehat{p}_{i}+A_{i}\right) ,  \label{H}
\end{equation}%
\noindent therefore, both have simultaneous eigenfunctions. It is important to mention that the value of
$l$ determines the direction the fermions takes along the axis, i.e., when $%
l>0$ the fermions go up and when $l<0$, they go down. Using the following ansatz%
\begin{equation}
\Psi ^{\pm }(u,v)=\frac{\phi^{\pm }(u)}{\sqrt[4]{f(u)}}\mathrm{e}^{i\omega lv},  \label{7}
\end{equation}%
\noindent the equations (\ref{6.1}) and (\ref{6.2}) can be reduced to
\begin{eqnarray}
i\left( \partial _{u}+\frac{l\omega }{\sqrt{f(u)}}-A_{v}\right) \phi ^{-} &=&-E\phi ^{+},  \label{8.1} \\
i\left( \partial _{u}-\frac{l\omega }{\sqrt{f(u)}}+A_{v}\right) \phi ^{+} &=&-E\phi ^{-}.  \label{8.2}
\end{eqnarray}%
\noindent These two coupled first-order equations can be decoupled for $E\neq 0$. By using the expression for $\phi ^{+}$ obtained from (\ref{8.1}) and inserting it in (\ref{8.2}) one obtains a second-order
differential equation for $\phi ^{-}$. In a similar way, using the
expression for $\phi ^{-}~$obtained from (\ref{8.2}) and inserting it in (%
\ref{8.1}) one obtains a second-order differential equation for $\phi ^{+}$.
It is possible to write both expressions as two supersymmetric
Schr\"{o}dinger-like equations%
\begin{equation}\label{9.1}
-\frac{d^{2}\phi ^{\pm}}{du^{2}}+V_{l}^{\pm}(u)\phi ^{\pm} =E^{2}\phi ^{\pm}\,,
\end{equation}
\noindent where the corresponding effective potentials are%
\begin{equation}\label{10.1}
V_{l}^{\pm}(u) =W_{l}^{2}\pm\frac{dW_{l}}{du},
\end{equation}
\noindent with the superpotential given by
\begin{equation}
W_{l}(u)=\frac{l\omega }{\sqrt{f(u)}}-A_{v}\,.  \label{11}
\end{equation}
These last results show that the solution for this class of problem consists in searching for bound-state solutions for two Schr\"{o}dinger-like equations. It should not be forgotten, though, that the equations for $\phi^{+}$ or $\phi^{-}$ are not indeed independent because $E$ appears in both equations. Therefore, one has to search for bound-state solutions for both signals in (\ref{9.1}) with a common energy.

\subsection{Free massless fermions}
\label{sec3:1:1}

\begin{figure}[t]
\subfloat[]{\label{fig_free1}\includegraphics[scale=0.60]{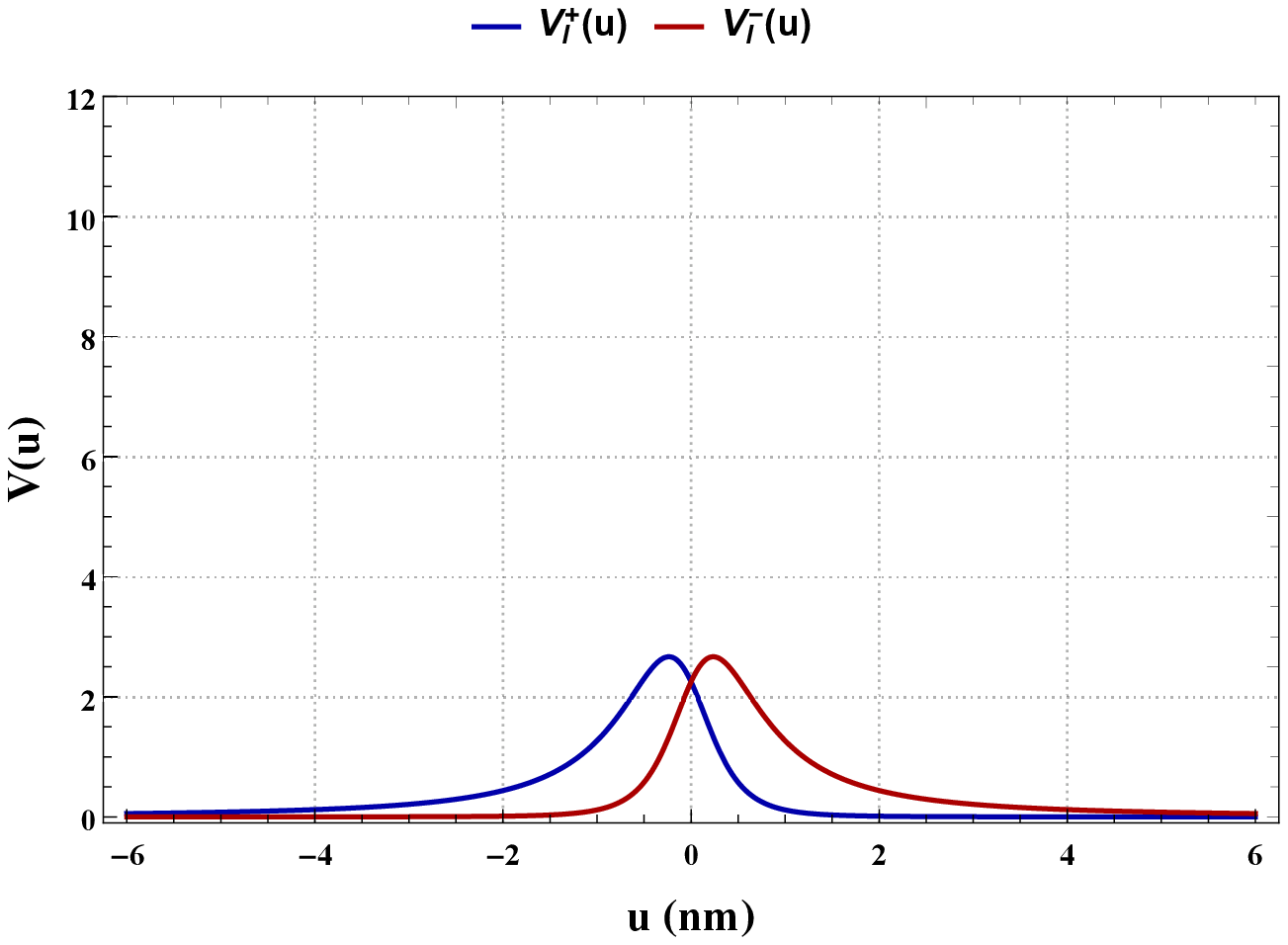}}\\
\subfloat[]{\label{fig_free2}\includegraphics[scale=0.60]{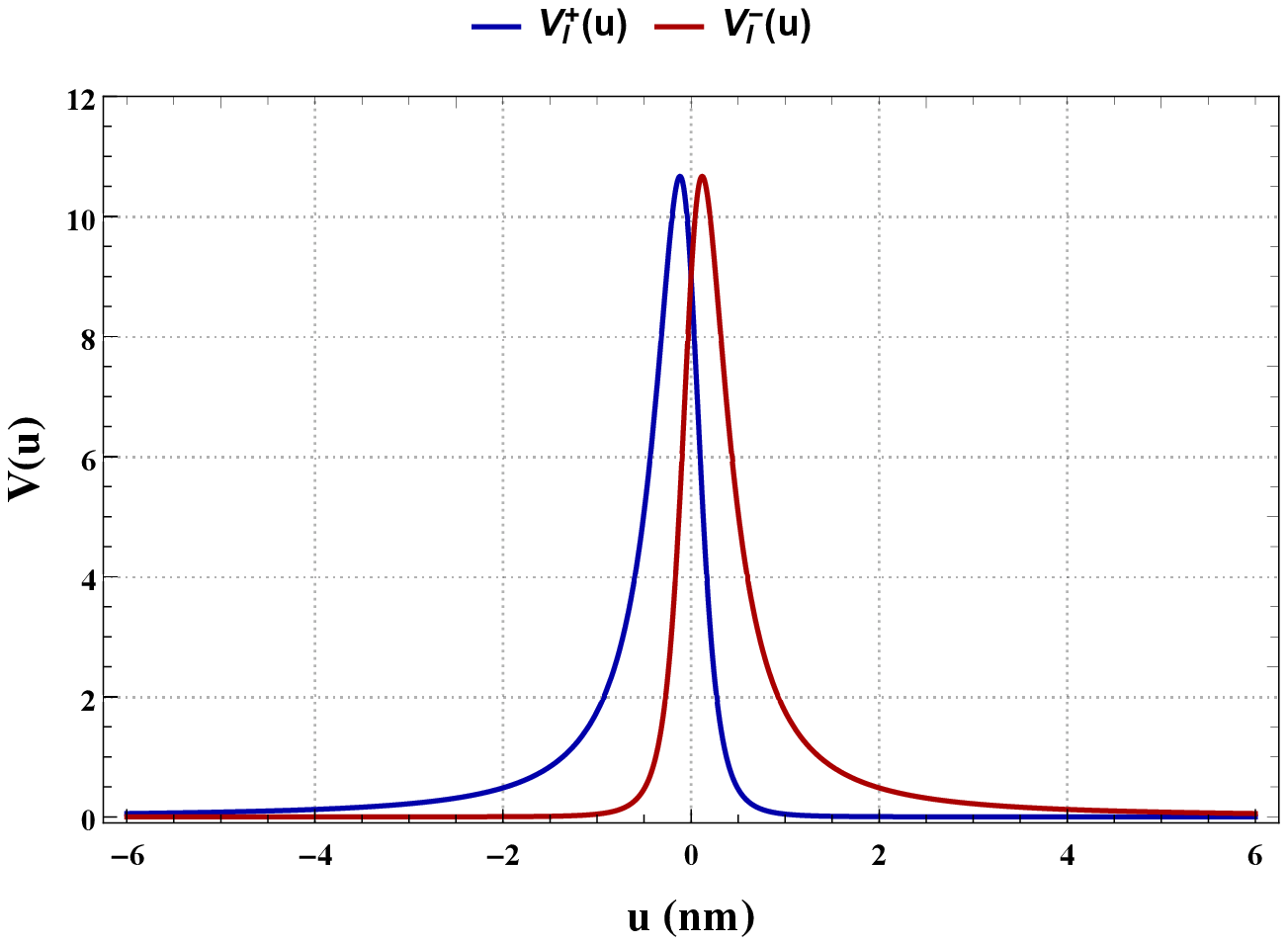}}
\caption{The panels show the effective potentials $V_{l}^{\pm }(u)$ for (a) $\protect\omega =1.5$ and (b) $\protect\omega =3.0$, considering $l=1$ and $B=0$ T.}
\label{fig_free}
\end{figure}

At this stage, we are interested in a vector potential $A_{\mu }=0$ ($B=0$). In this case, the two supersymmetric Schr\"{o}dinger-like equations (\ref{9.1}) and the corresponding effective potentials (\ref{10.1}) keep their mathematical structure intact with the superpotential given by
\begin{equation}\label{sp_free}
W_{l}(u)=\frac{l\omega }{\sqrt{f(u)}}\,.
\end{equation}
\noindent Substituting (\ref{sp_free}) into Eq.~(\ref{10.1}), we obtain
\begin{equation}\label{pot_ef_free}
V_{l}^{\pm }(u) = \left( \frac{l\omega }{\sqrt{f(u)}}\right) ^{2}\mp \left( \frac{u\omega ^{3}l}{\sqrt{f(u)}^{3}}\right)\,.
\end{equation}
\noindent It is important to highlight that this effective potential is caused by purely geometrical effects induced by the helicoidal parametrization. Note that from (\ref{pot_ef_free}) the effective potentials satisfy the relation $V_{l}^{+}(u)=V_{l}^{-}(-u)$. This result means that the change of chirality $\omega \rightarrow -\omega$ interchanges the effective potentials, thus $V_{l}^{+}(u)$ turns into $V_{l}^{-}(u)$, and viceversa. In figure \ref{fig_free}, we illustrate the behavior of the effective potential for $l=1$ and two different values of $\omega$, which emerges as the only control parameter of the system. In both cases, the profiles are composed by a barrier potential tending to zero as $u\rightarrow\pm\infty$ and there are only scattering states. Fixing $l=1$ in (\ref{pot_ef_free}), one can show that the peak of the barrier increases as $\omega$ increases. In this particular case, the maximum value of $V_{1}^{\pm}$ is $1.19\omega^{2}$.

\subsection{Uniform magnetic field}
\label{sec3:1}

\begin{figure}[b]
\begin{center}
\includegraphics[scale=1.1]{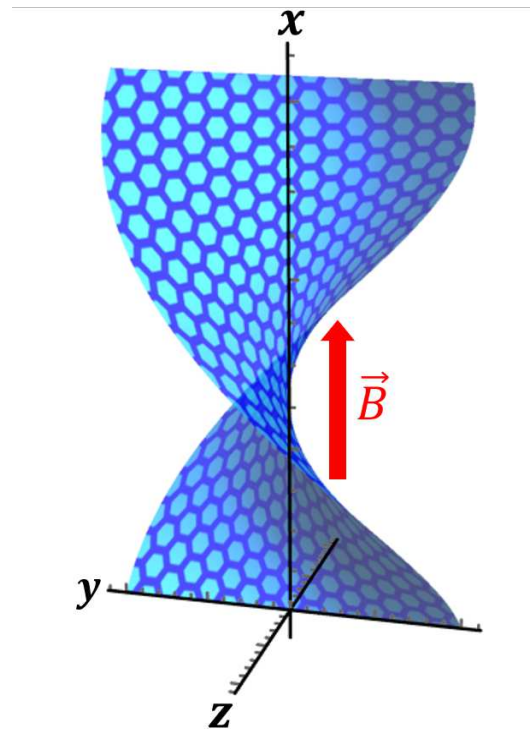}
\end{center}
\caption{Geometric structure of a helicoidal surface in
presence of a uniform magnetic $B$ in the direction
of its axis.}
\label{fig_1}
\end{figure}

At this stage, we are interested in a vector potential $A_{\mu }$ which provides a uniform
magnetic field aligned along the axis of helicoid, $\vec{B}=B\widehat{i}$ [see Fig.~(\ref{fig_1})], with $B>0$. This is obtained
considering the following configuration for the vector potential (see Appendix \ref{ap:A})%
\begin{equation}\label{12}
A_{v}(u)=\frac{B}{2\omega }\sqrt{f(u)}.
\end{equation}%
\noindent Substituting (\ref{12}) into Eq.~(\ref{10.1}), we have the following expression
for the effective potentials%
\begin{equation}\label{13}
\begin{split}
V_{l}^{\pm }(u) = & \left( \frac{l\omega }{\sqrt{f(u)}}-\frac{B}{2\omega }\sqrt{f(u)}\right) ^{2}\\
 & \mp \left( \frac{u\omega ^{3}l}{\sqrt{f(u)}^{3}}+\frac{u\omega B}{2\sqrt{f(u)}}\right)\,.
\end{split}
\end{equation}

\begin{figure}[t]
\begin{center}
\includegraphics[scale=0.60]{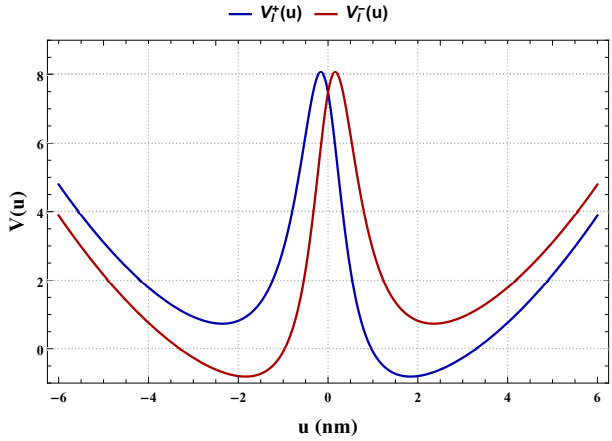}
\end{center}
\caption{Effective potentials for the helicoidal nanoribbon, for $l=2$, $w=1.5~$%
and $B=0.8$ T.}
\label{fig_2}
\end{figure}

\noindent From (\ref{13}), we can see that the effective potentials satisfy the relation $V_{l}^{+}(u)=V_{l}^{-}(-u)$. This last result means that the change of chirality $\omega \rightarrow -\omega$ interchanges the effective potentials, thus $V_{l}^{+}(u)$ turns into $V_{l}^{-}(u)$, and viceversa. The behavior of the effective potentials $V_{l}^{\pm }$ is plotted in figure \ref{fig_2} for $l=2$, $B=0.8$ T and $\omega=1.5$, where the value of the parameters related to the geometry of the nanoribbon were conveniently fixed at $D=12.0$~nm and $L=10\pi$. Figure \ref{fig_2} shows that the profiles of effective potentials $V_{l}^{\pm }$ are formed for two potential wells of different depth and a potential barrier between the wells. The presence of well structure is necessary for the existence of bound-state solutions and this is a consequence of the presence of the uniform magnetic field. Note that the two supersymmetric Schr\"{o}dinger-like equations (\ref{9.1}) with the effective potentials (\ref{13}) do not support exact solutions due to the chosen background geometry, so our results for the bound-state solutions and local density of state must be obtained numerically. This issue will be addressed in the following subsection.

\subsection{An application: Local density of states}
\label{sec3:2}

In this subsection, our goal is to analyze the way in which the states are distributed on the helicoidal surface by the effect of the uniform magnetic field. To achieve our goal, we need to calculate the local density of states (LDOS) of the system, which is defined by
\begin{equation}
\rho (u)=\frac{1}{2}\sum_{n,l}\left[ \left\vert \Psi
_{n,l}^{+}(u)\right\vert ^{2}+\left\vert \Psi _{n,l}^{-}(u)\right\vert ^{2}%
\right] ,  \label{14}
\end{equation}%
\noindent where $n$ is the principal quantum number.

\subsubsection{Free massless fermions}

\begin{figure}[t]
\subfloat[]{\label{fig_3a}\includegraphics[scale=0.60]{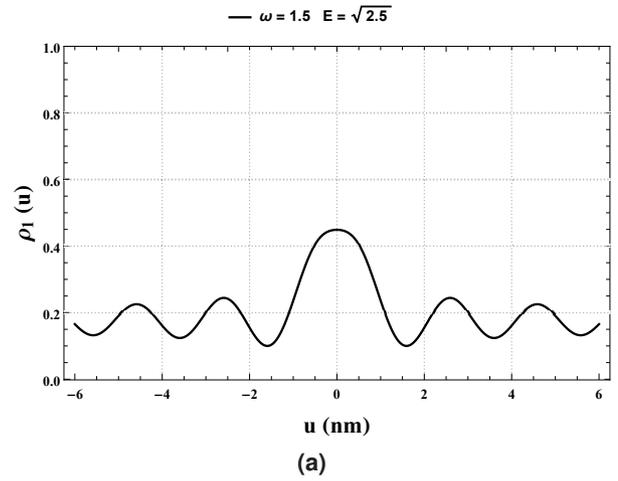}}\\
\subfloat[]{\label{fig_3b}\includegraphics[scale=0.60]{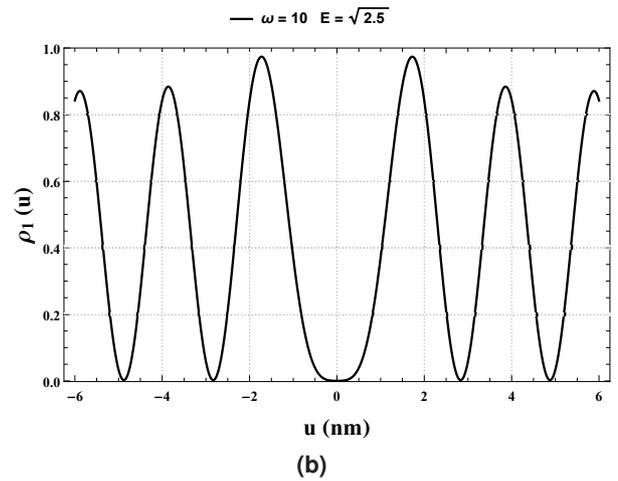}}
\caption{The panels show the partial local density of states $\protect\rho %
_{l}(u)~$for (a) $\protect\omega =1.5~$ and (b) $\protect\omega =10.0~$, considering $l=1$, $B=0$ T, and $E=\protect\sqrt{2.5}$.}
\label{fig_3}
\end{figure}

Figure \ref{fig_3} illustrates the behavior of the partial local density of states (PLDOS) $\rho _{l}(u)$ for $l=1$, $B=0$ T, $E=\sqrt{2.5}$, with $\omega =1.5~$ (Fig. \ref{fig_3a}) and $\omega =10.0~$(Fig. \ref{fig_3b}). As discussed in Sec.~(\ref{sec3:1:1}), the profiles of the effective potentials are composed by a barrier potential tending to zero as $u\rightarrow \pm \infty $. This profile implies that there are only scattering states. The maximum value of $V_{1}^{\pm }=1.19\omega ^{2}$, i.e, the peak of the barrier increases as $\omega$ increases. It is important to highlight that $\omega$ is also related to twist around to axis of helicoid (twist parameter), namely, $\omega =2\pi m/L$, where $m$ is the number of $2\pi$ twists. In our case, $\omega =1.5$ and $\omega =10.0$ are equivalent to $7.5$ and $50$ twisted, respectively. This mean that when the nanoribbon is twisted more times (large values of $\omega$), the states tend to shift for regions away from the axis of the helicoid. This behavior occurs due to increase in the intensity of the barrier (a more repulsive potential), which maintains its peak close to the axis of the helicoid. This analysis is in accordance with the results shown in the panels of Figure \ref{fig_3}, where we see that for $\omega =1.5~$(Fig. \ref{fig_3a}), the PLDOS is concentrated mainly in regions close to the axis of the helicoid. However, when $\omega ~$increases to $10.0$ (Fig. \ref{fig_3b}), we see that the PLDOS tends to zero in the axis of the helicoid, while the concentration of states increase in regions away from it. In Ref.~\cite{6}, the authors have investigated massless Dirac particles on a helicoid via the massless Dirac equation on curved space-time. They have showed that bound-states solutions are absent, and thus they have studied the scattering probabilities and the phase shifts. By means of numerical calculations, they have examined the LDOS around the axis of the helicoid. In \cite{6}, the authors do not consider a uniform magnetic field aligned along the axis of helicoid, nor the effects of the twist parameter on the effective potential profile, nor the effects of the twist parameter on the LDOS. Our results represent an extension to those shown in \cite{6}, where another parametrization is used.

\subsubsection{Uniform magnetic field}

\begin{figure}[t]
\subfloat[]{\label{fig_4a}\includegraphics[scale=0.59]{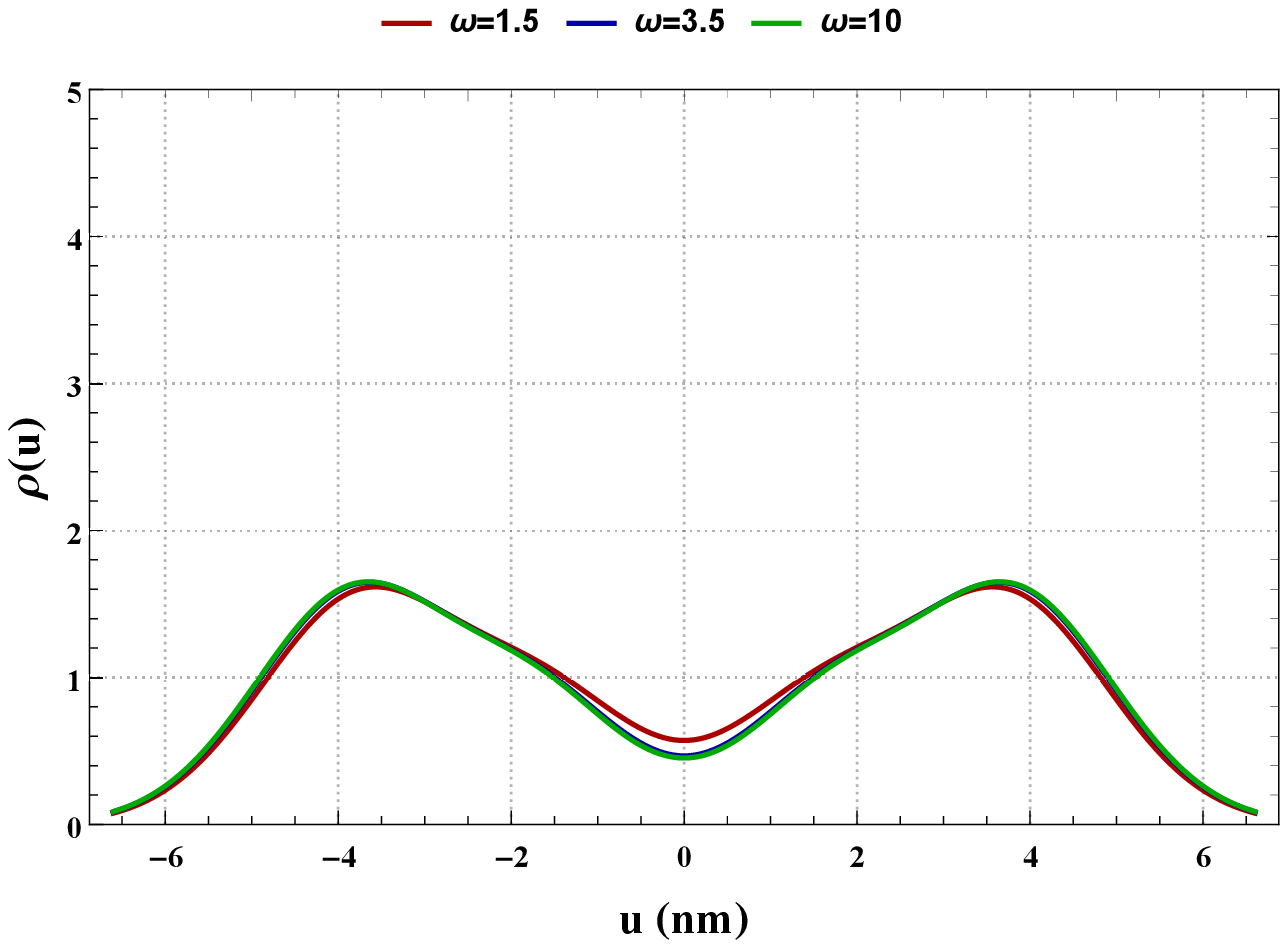}}\\
\subfloat[]{\label{fig_4b}\includegraphics[scale=0.59]{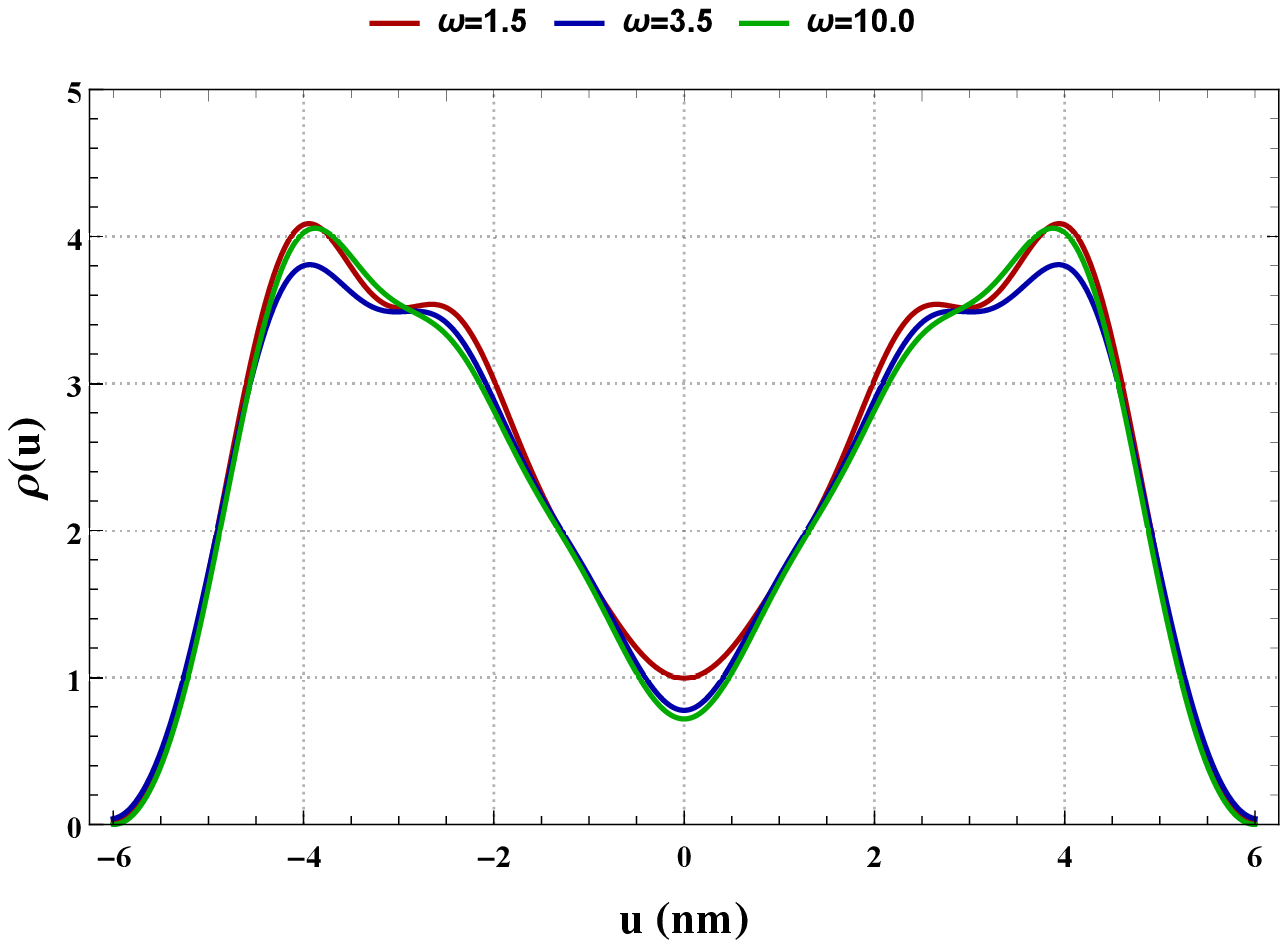}}
\caption{The panels show the local density of states $\protect\rho(u)~$for (a) $B=0.5$ T and (b) $B=0.8$ T, considering $D=12.0$~nm, $\protect\omega =0.5~$(black line), $\protect\omega =3.5~$%
(blue line), and $\protect\omega =10.0~$(red line).}
\label{fig_4}
\end{figure}

The results for $B\neq 0$ T are shown in the Figure \ref{fig_4}, where we illustrate the profiles of local density of states (LDOS) $\rho (u)$ for two values of magnetic field, $B=0.5$ T (Fig. \ref{fig_4a}) and $B=0.8$ T (Fig. \ref{fig_4b}). All bound states involved in the construction of $\rho (u)$ given by (\ref{14}) were computed numerically using the Numerov method. From (\ref{13}) we see that the height of the barrier between the wells increases as $\omega$ increases, while its width decreases as $B$ increases. From figure \ref{fig_4}, we observe the same behavior as in the case $B=0$ T, i.e., as $\omega$ increases, the tendency of the states is to concentrate on regions away from the axis of the helicoid. However, surprisingly, we see that for values of $\omega$ large enough, as for instance $\omega =10$, the LDOS on the axis is fixed at a single constant value. This point of minimum density of states is approximately $\rho _{\min }\approx 0.5$ for $B=0.5$ T and, $\rho _{\min }\approx 0.71$ for $B=0.8$ T. In this sense, we can infer that this peculiar behavior of the system is an effect of the confinement produced by the configuration of the uniform magnetic field, which maintains some states of particles captured to axis of the helicoid, no matter how many times it twists.

\begin{figure}[t]
\subfloat[]{\label{fig_5a}\includegraphics[scale=0.59]{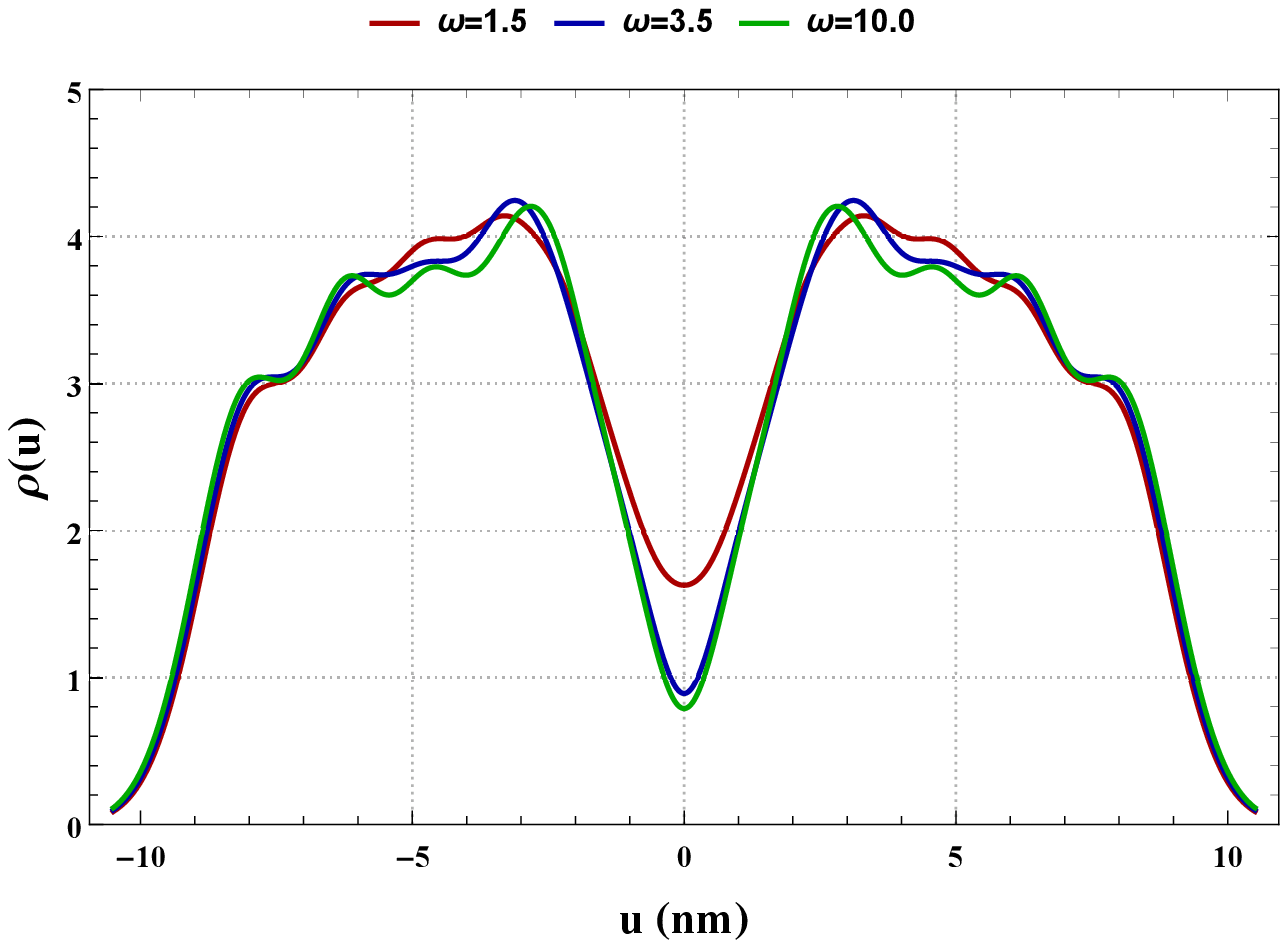}}\\
\subfloat[]{\label{fig_5b}\includegraphics[scale=0.59]{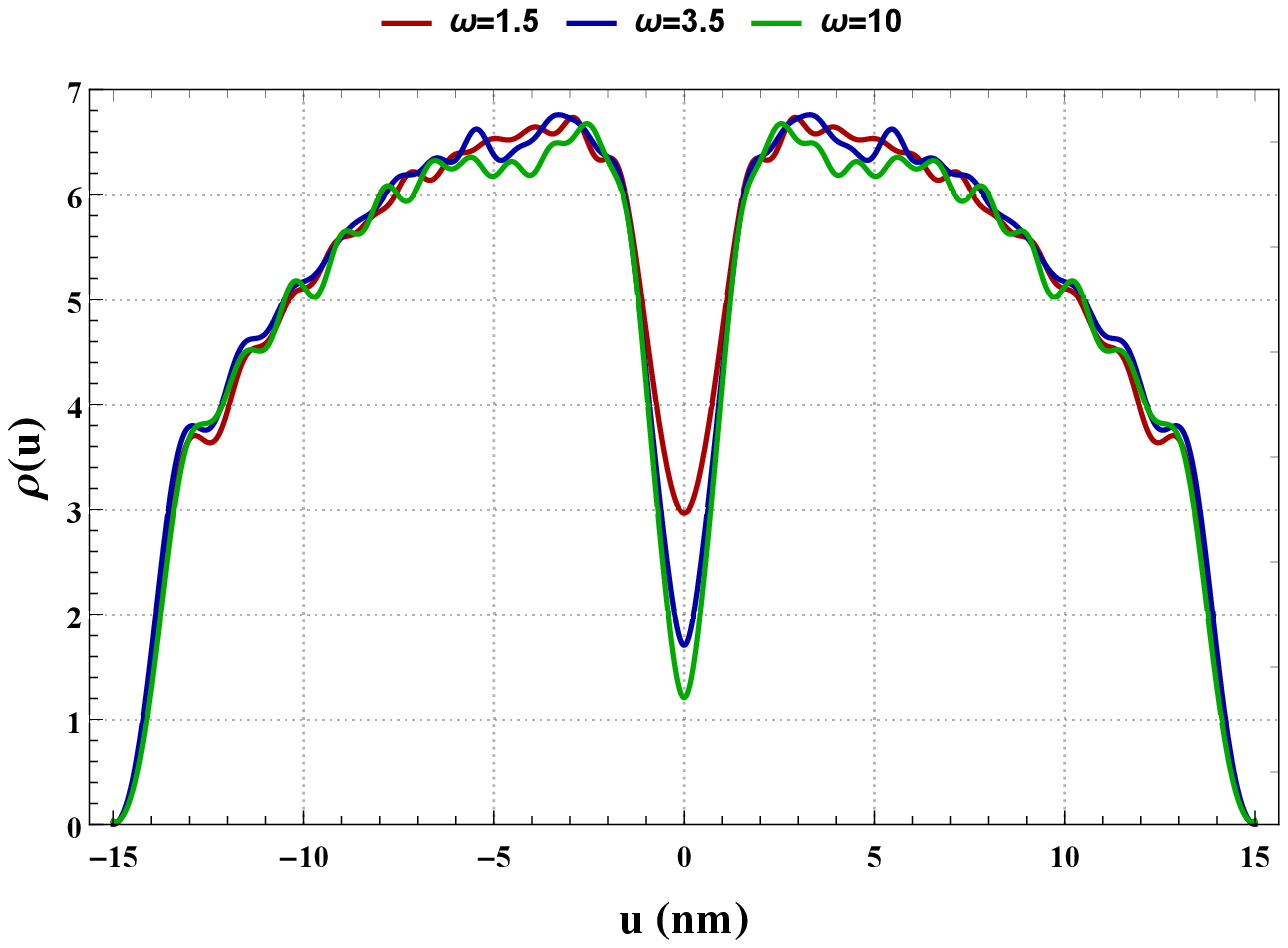}}
\caption{The panels show the local density of states $\rho(u)$ for (a) $D=20.0$~nm and (b) $D=30.0$~nm, considering $B=0.5$~T, $\omega=0.5$ (black line), $\omega=3.5$ (blue line), and $\omega=10.0$ (red line).}
\label{fig_5}
\end{figure}

\noindent This is a surprising and highly counterintuitive result, which occurs from $\omega \approx 3.5$ onwards, according to Figure \ref{fig_4}. From a qualitative perspective, we believe that this particular value of $\omega $ could be related to the so-called \textquotedblleft critical angle\textquotedblright, which determines the breaking point or fracture of the flat elastic nanoribbon subject to twist \cite{chopin2013helicoids,1}. Assuming that this fracture of the helicoidal nanoribbon occurs at points on its axis (the lattice constant is approximated locally to be zero) \cite{1}, it is reasonable to assume that the density of states at $u=0$ is minimized. According to this, the critical angle for the free case ($B=0~$T) would occur for $\omega_{c}\approx 10$, when the states leave the axis of the helicoid ($\rho_{l}\rightarrow 0$) due to a possible fracture on the axis. However, when a uniform magnetic field is introduced into the system, this critical angle decreases to $\omega _{c}\approx 3.5$, that is, the magnetic field accelerates the process of breaking the graphene, but keeping some states trapped on the axis ($\rho\rightarrow cte$). By the other hand, in \cite{1} was showed (via DFT calculations) that a isolated graphene lattice reconstructs itself after twisting beyond a critical angle. This behavior is typically associated with some kind of phase transition observed in a twist energy diagram, where the discontinuity in the transition zone occurs exactly at the critical angle. Although the methods used in this work do not allow us to analyze such phase transitions, it is possible to appreciate some indications of such discontinuities for large values of $\omega $ (strongly twisted nanoribbon), specifically, in our effective potentials, which becomes
\begin{equation}
V_{l}^{+}(u) = \frac{(Bu-2l)^2}{4u^{2}}, \\
V_{l}^{-}(u) = \frac{Bu^{2}+2l}{2u^{2}}.
\end{equation}%
Note that the above expressions are singular in $u=0$, and, furthermore, their solutions will also be singular at the same point. In this sense, we can say that for values of $\omega<\omega_{c}$, the system remains in one stable ordered phase, where the predominant concentration of states occurs in the region near to the axis of the nanoribbon. However, as we approach the critical value $\omega_{c}$, the LDOS adopts a constant value at the axis, which is an effect produced by those states trapped by the magnetic field. Finally, when $\omega=\omega_{c}$, the local density of states becomes discontinuous due to the singular states, which means that a phase transition is about to occur. In Ref.~\cite{17}, the authors have considered the quantum mechanics of a charged particle on a helicoid in an external magnetic field via the Schrödinger equation on a two-dimensional curved surface. Choosing two simple magnetic field configurations, they have examined the behavior of the effective potentials for different values of angular momentum and the applied magnetic field strength. They also have obtained approximate expressions for the energy levels, which are valid when the particle is near a minimum and these are similar to the energy levels of a particle in a harmonic oscillator potential. In \cite{17}, the authors do not consider the massless Dirac equation on curved space-time, nor the effects of twist parameter on the effective potential profile, nor the construction of the LDOS, nor a uniform magnetic field aligned along the axis of their infinite helicoid. With respect to this last point, it is important to mention that a infinite helicoidal graphene ($D$ and $L\rightarrow \infty$) represents non-realizable scenario in the context of our parametrization. This is because the twist parameter $\omega=2\pi m/L$ tends to zero as $L\rightarrow \infty$, which transforms the helicoidal geometry of graphene to its flat version (see Figure \ref{fig_w}). By the other hand, a semi-infinite helicoidal graphene (fixed $L$ and $D\rightarrow \infty$) provides a more realistic scenario, nevertheless, we have $\rho_{min}\rightarrow \infty$, due to the infinite states confined within the infinite potential well. This means that it is not possible to determine a break point of the material (there will be no phase transition), and any control parameters related to the twist of helicoidal graphene will be irrelevant. These statements are verified from the Figure \ref{fig_5}, which shows the behavior of the LDOS for different values of $D$, where it can be seen that $\rho_{min}$ increases with increasing $D$. In this way, when $D \rightarrow \infty$, we have $\rho_{min}\rightarrow \infty$.


\section{Conclusions}
\label{sec:4}

We have studied the problem of massless Dirac fermions moving in (2+1)-dimensional on a helicoidal surface in the presence of a uniform magnetic field aligned along the axis of the helicoidal nanoribbon. We have shown that the standard decomposition of this Dirac equation into its upper and lower components produces a set of two supersymmetric Schr\"{o}dinger-like equations. The effective potential for each component ($B\neq 0$ T) is formed by two potential wells separated by a finite barrier, which is completely confining due to the presence of magnetic field. Nevertheless, the number of bound states is finite because the dimensions of the helicoidal nanoribbon are also finite. On the other hand, for the free case ($B=0$ T), where only scattering states are present, the effective potential is a barrier potential, which, for small values of $\omega$ (slightly twisted nanoribbon), produces a LDOS whose region of predominant concentration occurs at points near to the axis of the nanoribbon. However, when $\omega$ increases, the repulsive intensity of the barrier also increases, making the concentration of states to shift in regions away from the axis, where the partial density of states tends to zero. It was observed that the behavior of the LDOS for $B\neq 0$ T is the same as in the free case $B=0$ T for similar values of $\omega$. Nevertheless, for values large enough of $\omega$ (strongly twisted nanoribbon), the LDOS on the axis reaches a constant value of minimum density, i.e., that no matter how many times the helicoid is twisted, there will always be a concentration of states on its axis. This novel result is produced by the configuration of the uniform magnetic field.

As mentioned, a feasible experimental realization of our results can be occurs in the context of condensed matter Physics, particularly, in the study of the phase transitions where the twist angle (associated to our $\omega$), is used as a control parameter, which determines the breaking point of a nanoribbon subject to twist. Although this connection with the critical angle and the phase transitions was carried out from a qualitative approach, we believe that the results found in this work will be useful to extend those shown in \cite{chopin2013helicoids,1}. In fact, this analysis is currently in development and will be shown in our next work.

\begin{acknowledgements}
The authors are indebted to the anonymous referees for an excellent and constructive review. This work was supported in part by means of funds provided by CNPq, Brazil, Grant No. 311925/2020-0 (PQ) and Grant No. 422755/2018-4 (UNIVERSAL), FAPEMA, Brazil, Grant No. UNIVERSAL-01220/18, and CAPES, Brazil.
\end{acknowledgements}

\appendix	

\section{Derivation of the electromagnetic potential}
\label{ap:A}

Using the following parametrization
\begin{equation}\label{f1}
\PC{
\begin{array}{c}
x \\
y \\
z
\end{array}
}  =
\PC{
\begin{array}{c}
v \\
u \cos \PC{\omega v} \\
u \sin \PC{\omega v}
\end{array}
},
\end{equation}
\noindent the position vector $\vec{r}$ is given by
\begin{equation}\label{f2}
\vec{r} = v \hat{i} + u \cos \PC{\omega v} \hat{j} + u \sin \PC{\omega v} \hat{k}.
\end{equation}
\noindent
Performing the standard procedure, one can write the new bases as
\begin{equation}\label{f3}
\hat{e}_{i}  =   \frac{\partial_{i} \vec{r} \PC{u , v}}{|\partial_{i} \vec{r} \PC{u , v}|}, \quad \mathrm{for} \quad i=u,v\,.
\end{equation}
From this last expression, we get
\begin{subequations}
\begin{equation}\label{f4}
\hat{e}_{u} = \cos \PC{\omega v} \hat{j} +  \sin \PC{\omega v} \hat{k} \,,
\end{equation}
\begin{equation}\label{f5}
\hat{e}_{v} = \frac{ \hat{i} - \omega u \sin \PC{\omega v} \hat{j} + \omega u \cos \PC{\omega v} \hat{k}}{\sqrt{1 +{u}^{2} {\omega}^{2}}}\,,
\end{equation}
\end{subequations}
\noindent where was used
\begin{subequations}
\begin{equation}\label{f6}
\partial_{u} \vec{r} \PC{u , v} = \cos \PC{\omega v} \hat{j} +  \sin \PC{\omega v} \hat{k}\,,
\end{equation}
\begin{equation}\label{f7}
\partial_{v} \vec{r} \PC{u , v} = \hat{i} - \omega u \sin \PC{\omega v}  \hat{j} + \omega u \cos \PC{\omega v} \hat{k}\,.
\end{equation}
\end{subequations}
\noindent At this point, one can do the inverse procedure to write the Cartesian partial derivates ($\partial_{x}, \partial_{y}, \partial_{z}$) as a function of the partial derivates $\partial_{u}$ and $\partial_{v}$. In this way, we have\begin{subequations}
\begin{equation}\label{f8}
\partial_{x} = \frac{\partial }{\partial v}\,,
\end{equation}
\begin{equation}\label{f9}
\partial_{y} = \cos \PC{\omega v} \frac{\partial}{\partial u}  -  \frac{\sin \PC{\omega v}}{{\omega u}} \frac{\partial}{\partial v}\,,
\end{equation}
\begin{equation}\label{f10}
\partial_{z} = \sin \PC{\omega v} \frac{\partial}{\partial u}  +  \frac{\cos \PC{\omega v}}{{\omega u}} \frac{\partial}{\partial v}.
\end{equation}
\end{subequations}
\noindent Then, using the bases $\hat{e}_{u}$ and $\hat{e}_{v}$, one can find the spatial component of the electromagnetic four-potential $A^{\alpha} = \PC{A_{0}, \vec{A}}$, that is
\begin{equation}
\vec{A} = A_{u} \hat{e}_{u} + A_{v} \hat{e}_{v}=A_{x}\hat{i}+A_{y}\hat{j}+A_{z}\hat{k}\,,
\end{equation}
\noindent where
\begin{subequations}
\begin{equation}\label{f12}
A_{x} = \frac{A_{v}}{\sqrt{1 +{u}^{2} {\omega}^{2}}}
\end{equation}
\begin{equation}\label{f13}
A_{y} = A_{u}\cos \PC{\omega v} - A_{v} \frac{\omega u \sin \PC{\omega v}}{\sqrt{1 +{u}^{2} {\omega}^{2}}}
\end{equation}
\begin{equation}\label{f14}
A_{z} = A_{u} \sin \PC{\omega v} + A_{v} \frac{\omega u \cos \PC{\omega v}}{\sqrt{1 +{u}^{2} {\omega}^{2}}}.
\end{equation}
\end{subequations}

\subsection{Uniform magnetic field}\label{subsec4.1.2}

One can calculate the magnetic field aligned along the axis of the helicoid ($\vec{B} = B \hat{i}$) from
\begin{equation}\label{f15}
B = \partial_{y} A_{z} - \partial_{z} A_{y}.
\end{equation}
\noindent Substituting the expressions \eqref{f9}, \eqref{f10}, \eqref{f13} and \eqref{f14}, we get
\begin{equation}\label{f16}
B = \frac{\partial}{\partial u}\PC{\frac{\omega u A_{v}}{\sqrt{1 +{u}^{2} {\omega}^{2}}}} + \frac{\omega A_{v}}{\sqrt{1 +{u}^{2} {\omega}^{2}}} - \frac{1}{\omega u}\frac{\partial}{\partial v} A_{u}.
\end{equation}
\noindent In order to obtain a uniform magnetic field ($B = \mathrm{cte}$), one can properly choose $A_{u} = 0$ and
\begin{equation}\label{f18}
A_{v} = \frac{B}{2 \omega} \sqrt{1 +{u}^{2} {\omega}^{2}}.
\end{equation}

\bibliographystyle{andp2012}
\providecommand{\WileyBibTextsc}{}
\let\textsc\WileyBibTextsc
\providecommand{\othercit}{}
\providecommand{\jr}[1]{#1}
\providecommand{\etal}{~et~al.}

\end{document}